\documentclass{article}
\usepackage{amsfonts}

\usepackage{amsmath}
\usepackage{cite}
\usepackage{graphicx}
\usepackage{dcolumn}

\renewcommand{\Bbb}{\mathbb}

\begin{document}
\date{}
\title{On the normal modes of coupled harmonic oscillators}
\author{Francisco M. Fern\'{a}ndez\thanks{%
fernande@quimica.unlp.edu.ar} \\
INIFTA, DQT, Sucursal 4, C. C. 16, \\
1900 La Plata, Argentina} \maketitle
\begin{abstract}
We discuss in detail a well known method for obtaining the
frequencies of the normal modes of coupled harmonic oscillators
that is based on the simultaneous diagonalization of two symmetric
matrices. We apply it to some simple illustrative examples
recently chosen for the presentation of an alternative approach
based on a set of constants of the motion. We show that the
traditional method is simpler, even in the case of equal
frequencies. We also discuss a problem that commonly appears in
elementary courses on quantum mechanics that also requires the
diagonalization of two symmetric matrices.
\end{abstract}

\section{Introduction}

\label{sec:intro}

The calculation of the normal modes of coupled harmonic oscillators is
commonly discussed in most textbooks on classical mechanics\cite{G80} and is
of relevance in the analysis of the vibrational spectroscopy of polyatomic
molecules in terms of internal nuclear coordinates\cite{WDC55}. The \textit{%
traditional} treatment of the problem is based on the simultaneous
diagonalization of two symmetric matrices for the kinetic an
potential energies\cite{WDC55,C78}. This approach also applies to
the quantum-mechanical version of the problem\cite{F20b}. In a
recent paper, Hojman\cite{H23} proposed an alternative method for
obtaining the frequencies of the normal modes based on a set of
constants of the motion. In our opinion it may be of great
pedagogical interest to compare both approaches.

In section~\ref{sec:CHO} we develop the traditional method in
detail and apply it to the models proposed by Hojman in
section~\ref{sec:examples}, where we also discuss an additional
three-particle problem with degenerate eigenfrequencies. In
appendix~\ref{sec:analogy} we outline a problem that commonly
appears in elementary courses on quantum mechanics and also
requires the simultaneous diagonalization of two symmetric
matrices. Finally, in section~\ref{sec:conclusions} we summarize
the main results and draw conclusions.

\section{Diagonalization of coupled harmonic oscillators}

\label{sec:CHO}

In this paper we consider $N$ coupled harmonic oscillators with kinetic
energy $T$ and potential energy $V$ given by
\begin{equation}
T=\frac{1}{2}\mathbf{\dot{Q}}^{t}\mathbf{A\dot{Q},\;V=}\frac{1}{2}\mathbf{Q}%
^{t}\mathbf{BQ},  \label{eq:T,V}
\end{equation}
where $\mathbf{Q}$ and $\mathbf{\dot{Q}}$ are column vectors for the
coordinates $q_{i}$ and velocities $\dot{q}_{i}$, $i=1,2,\ldots ,N$,
respectively. The superscript $t$ stands for transpose and $\mathbf{A}$ and $%
\mathbf{B}$ are time-independent $N\times N$ symmetric matrices. From a
physical point of view, we assume that $\mathbf{A}$ is positive definite
(all its eigenvalues are positive real numbers).

In order to bring both $\mathbf{A}$ and $\mathbf{B}$ into diagonal form we
propose the change of variables $\mathbf{Q}=\mathbf{CS}$, where $\mathbf{C}$
is an $N\times N$ invertible matrix and $\mathbf{S}$ a column vector for the
new coordinates $s_{i}$, $i=1,2,\ldots ,N$. We choose $\mathbf{C}$ to
satisfy two conditions, first
\begin{equation}
\mathbf{C}^{t}\mathbf{BC}=\mathbf{\Lambda },  \label{eq:CBC=Lambda}
\end{equation}
where $\mathbf{\Lambda }=\left( \lambda _{i}\delta _{ij}\right) $ is a
diagonal matrix, second
\begin{equation}
\mathbf{C}^{t}\mathbf{AC}=\mathbf{I},  \label{eq:CAC=I}
\end{equation}
where $\mathbf{I}$ is the $N\times N$ identity matrix. We will show below
that this matrix already exists. It follows from equation (\ref{eq:CAC=I})
that $\mathbf{C}^{t}=\mathbf{C}^{-1}\mathbf{A}^{-1}$ and equation (\ref
{eq:CBC=Lambda}) thus becomes
\begin{equation}
\mathbf{C}^{-1}\mathbf{A}^{-1}\mathbf{BC}=\mathbf{\Lambda }.  \label{eq:CABC}
\end{equation}
This equation simply represents the diagonalization of the matrix $\mathbf{A}%
^{-1}\mathbf{B}$. More precisely, the problem reduces to obtaining the
eigenvalues $\lambda _{i}$ and the column eigenvectors $\mathbf{C}_{i}$ of
the matrix $\mathbf{A}^{-1}\mathbf{B}$. The columns of the matrix $\mathbf{C}
$ are precisely such eigenvectors. Since the eigenvectors $\mathbf{C}_{i}$
are not normalized we use equation (\ref{eq:CAC=I}) to obtain their norms.
Finally, the physical problem reduces to solving the trivial equations of
motion for a set of $N$ uncoupled harmonic oscillators:
\begin{equation}
T=\frac{1}{2}\mathbf{\dot{S}}^{t}\mathbf{S}=\frac{1}{2}\sum_{i=1}^{N}\dot{s}%
_{i}^{2},\;V=\frac{1}{2}\mathbf{S}^{t}\mathbf{\Lambda S=}\frac{1}{2}%
\sum_{i=1}^{N}\lambda _{i}s_{i}^{2}.  \label{eq:T,V_diag}
\end{equation}
It only remains to prove that $\mathbf{A}^{-1}\mathbf{B}$ is diagonalizable.

Since $\mathbf{A}$ is positive definite, then $\mathbf{A}^{1/2}$ exists and
we can construct the new matrix $\mathbf{U}=\mathbf{A}^{1/2}\mathbf{C}$ that
is orthonormal ($\mathbf{U}^{t}=\mathbf{U}^{-1}$) as shown by
\begin{equation}
\mathbf{U}^{t}\mathbf{U}=\mathbf{C}^{t}\mathbf{AC}=\mathbf{I}.  \label{eq:UU}
\end{equation}
If we substitute $\mathbf{C=A}^{-1/2}\mathbf{U}$ into equation (\ref
{eq:CBC=Lambda}) we obtain
\begin{equation}
\mathbf{U}^{t}\mathbf{A}^{-1/2}\mathbf{BA}^{-1/2}\mathbf{U}=\mathbf{\Lambda }%
\text{.}  \label{eq:UABAU=Lambda}
\end{equation}
Since $\mathbf{A}^{-1/2}\mathbf{BA}^{-1/2}$ is symmetric, then it is
diagonalizable, the orthogonal matrix $\mathbf{U}$ exists\cite{GV96} and,
consequently, the matrix $\mathbf{C}$ also exists. It is clear that we can
always diagonalize the kinetic and potential energies for a system of
coupled harmonic oscillators as shown in equation (\ref{eq:T,V_diag}).
Present analysis of the problem posed by the simultaneous diagonalization of
two symmetric matrices appears to be simpler than the one proposed by Chavda%
\cite{C78} some time ago.

Before discussing suitable illustrative examples, it is convenient to pay
attention to the units of the matrices introduced above. The matrix elements
of $\mathbf{A}$ and $\mathbf{B}$ have units of mass and energy$\times $length%
$^{-2}$, respectively. Consequently, the elements of $\mathbf{C}$ have units
of mass$^{-1/2}$ and the new variables $s_{i}$ have units of mass$%
^{1/2}\times $length. Finally, the eigenvalues $\lambda _{i}$ have units of
time$^{-2}$. If we write $\lambda _{i}=\omega _{i}^{2}$, then $\omega _{i}$,
$i=1,2,\ldots ,N$, are the frequencies of the normal modes. The new
variables $s_{i}$ can be interpreted as a kind of mass-weighted coordinates
for the normal modes.

\section{Examples}

\label{sec:examples}

The first example is the two-dimensional model chosen by Hojman\cite{H23}
\begin{equation}
T=\frac{m}{2}\left( \dot{q}_{1}^{2}+\dot{q}_{2}^{2}\right) ,\;V=\frac{1}{2}%
\left[ k_{1}\left( q_{1}^{2}+q_{2}^{2}\right) +k_{2}\left(
q_{1}-q_{2}\right) ^{2}\right] ,  \label{eq:T,V_2D}
\end{equation}
that leads to
\begin{equation}
\mathbf{A}=\Bbb{T}_{2}=m\mathbf{I},\;\mathbf{B}=\Bbb{V}_{2}=\left(
\begin{array}{rr}
k_{1}+k_{2} & -k_{2} \\
-k_{2} & k_{1}+k_{2}
\end{array}
\right) ,  \label{eq:A,B_2D}
\end{equation}
and
\begin{equation}
\mathbf{A}^{-1/2}\mathbf{BA}^{-1/2}=\mathbf{A}^{-1}\mathbf{B}=\frac{1}{m}%
\mathbf{B}.  \label{eq:ABA_2D}
\end{equation}
The two eigenvalues of this matrix are
\begin{equation}
\lambda _{1}=\frac{k_{1}}{m},\;\lambda _{2}=\frac{k_{1}+2k_{2}}{m},
\label{eq:lambda_i_2D}
\end{equation}
with eigenvectors
\begin{equation}
\mathbf{U}_{1}=\alpha \left(
\begin{array}{r}
1 \\
1
\end{array}
\right) ,\;\mathbf{U}_{2}=\beta \left(
\begin{array}{r}
1 \\
-1
\end{array}
\right) .  \label{eq:C_i_2D}
\end{equation}
They are orthogonal and we can choose $\alpha =\beta =1/\sqrt{2}$ as
normalization factors. Therefore, the matrices $\mathbf{U}$ and $\mathbf{C}$
are given by
\begin{equation}
\mathbf{U}=\frac{1}{\sqrt{2}}\left(
\begin{array}{rr}
1 & 1 \\
1 & -1
\end{array}
\right) ,\;\mathbf{C}=\frac{1}{\sqrt{m}}\mathbf{U}.  \label{eq:U,C_2D}
\end{equation}
The resulting normal modes
\begin{equation}
\mathbf{U}\left(
\begin{array}{l}
q_{1} \\
q_{2}
\end{array}
\right) =\frac{1}{\sqrt{2}}\left(
\begin{array}{l}
q_{1}+q_{2} \\
q_{1}-q_{2}
\end{array}
\right) ,  \label{eq:UQ_2D}
\end{equation}
agree with the ones derived by Hojman\cite{H23}.

As a three-dimensional example, Hojman chose the toy model
\begin{equation}
\mathbf{A}=\Bbb{T}_{3}=\left(
\begin{array}{rrr}
9 & -23 & -22 \\
-23 & 61 & 58 \\
-22 & 58 & 56
\end{array}
\right) ,\;\mathbf{B}=\Bbb{V}_{3}=\left(
\begin{array}{rrr}
15 & -37 & -34 \\
-37 & 95 & 86 \\
-34 & 86 & 80
\end{array}
\right) .  \label{eq:A,B_3D_toy}
\end{equation}
It is not difficult to verify that the matrix $\mathbf{A}$ is
positive definite (its three eigenvalues are positive). The
eigenvalues and eigenfunctions of
\begin{equation}
\mathbf{A}^{-1}\mathbf{B}=\left(
\begin{array}{rrr}
4 & -6 & -6 \\
-1 & 5 & 2 \\
2 & -6 & -3
\end{array}
\right) ,  \label{eq:AB_toy}
\end{equation}
are
\begin{eqnarray}
\lambda _{1} &=&1,\;\lambda _{2}=2,\;\lambda _{3}=3,  \nonumber \\
\mathbf{C}_{1} &=&\alpha _{1}\left(
\begin{array}{r}
1 \\
0 \\
\frac{1}{2}
\end{array}
\right) ,\;\mathbf{C}_{2}=\alpha _{2}\left(
\begin{array}{r}
1 \\
\frac{1}{3} \\
0
\end{array}
\right) ,\;\mathbf{C}_{3}=\alpha _{3}\left(
\begin{array}{r}
0 \\
1 \\
-1
\end{array}
\right) .  \label{eq:lambda_i,C_i_toy}
\end{eqnarray}
It is clear that the matrices $\mathbf{A}$ and $\mathbf{B}$ were purposely
chosen to have extremely simple results. It follows from equation (\ref
{eq:CAC=I}) that $\alpha _{1}=\pm 1$, $\alpha _{2}=\pm 3/2$, $\alpha
_{3}=\pm 1$. On arbitrarily selecting the positive signs, without loss of
generality, we have
\begin{equation}
\mathbf{C}=\left(
\begin{array}{rrr}
1 & \frac{3}{2} & 0 \\
0 & \frac{1}{2} & 1 \\
\frac{1}{2} & 0 & -1
\end{array}
\right) ,  \label{eq:C_toy}
\end{equation}
that satisfies both equations (\ref{eq:CAC=I}) and (\ref{eq:CBC=Lambda}) as
one can easily verify.

The mass-weighted coordinates
\begin{eqnarray}
-\frac{1}{2}s_{1} &=&q_{1}-3\left( q_{2}+q_{3}\right) ,  \nonumber \\
\frac{1}{2}s_{2} &=&q_{1}-2\left( q_{2}+q_{3}\right) ,  \nonumber \\
-\frac{1}{2}s_{3} &=&\frac{1}{2}\left( q_{1}-3q_{2}-2q_{3}\right) ,
\label{eq:s_i_toy}
\end{eqnarray}
agree with those obtained by Hojman\cite{H23} through a lengthier procedure
based on the construction of the constants of the motion.

In what follows, we discuss a somewhat more realistic three-dimensional
model given by a set of three identical particles with harmonic
interactions:
\begin{equation}
T=\frac{m}{2}\left( \dot{q}_{1}^{2}+\dot{q}_{2}^{2}+\dot{q}_{3}^{2}\right)
,\;V=\frac{k}{2}\left[ \left( q_{1}-q_{2}\right) ^{2}+\left(
q_{1}-q_{3}\right) ^{2}+\left( q_{2}-q_{3}\right) ^{2}\right] .
\label{eq:T,V_3D_identical}
\end{equation}
The relevant matrices are
\begin{equation}
\mathbf{A}=m\mathbf{I},\;\mathbf{B}=k\left(
\begin{array}{rrr}
2 & -1 & -1 \\
-1 & 2 & -1 \\
-1 & -1 & 2
\end{array}
\right) .  \label{eq:A,B_3D_identical}
\end{equation}
The eigenvalues and eigenvectors of $\mathbf{A}^{-1/2}\mathbf{BA}%
^{-1/2}=m^{-1}\mathbf{B}$ are
\begin{eqnarray}
\lambda _{1} &=&0,\;\lambda _{2}=\lambda _{3}=\frac{3k}{m},  \nonumber \\
\mathbf{U}_{1} &=&\alpha _{1}\left(
\begin{array}{l}
1 \\
1 \\
1
\end{array}
\right) ,\;\mathbf{U}_{2}=\left(
\begin{array}{r}
\alpha _{2} \\
\alpha _{3} \\
-\alpha _{2}-\alpha _{3}
\end{array}
\right) ,\;U_{3}=\left(
\begin{array}{r}
\alpha _{4} \\
\alpha _{5} \\
-\alpha _{4}-\alpha _{5}
\end{array}
\right) .  \label{eq:lambda_i,C_i_3D_identical}
\end{eqnarray}
In the presentation of his method, Hojman\cite{H23} assumed, for simplicity,
that all the frequencies were different. In the present case, we appreciate
that the application of the traditional approach to the case of equal
frequencies is straightforward. The column vector $\mathbf{U}_{1}$ is
orthogonal to the other two and we normalize it by choosing $\alpha _{1}=1/%
\sqrt{3}$. Without loss of generality we arbitrarily choose $\alpha _{3}=0$
and $\alpha _{2}=1/\sqrt{2}$. From $\mathbf{U}_{2}^{t}\mathbf{U}_{3}=0$ we
obtain $\alpha _{5}=-2\alpha _{4}$ and $\mathbf{U}_{3}^{t}\mathbf{U}_{3}=1$
yields $\alpha _{4}=\pm 1/\sqrt{6}$. Finally, the matrix $\mathbf{C}$
becomes
\begin{equation}
\mathbf{C}=\frac{1}{\sqrt{m}}\mathbf{U},\;\mathbf{U}=\frac{1}{6}\left(
\begin{array}{rrr}
2\sqrt{3} & 3\sqrt{2} & \sqrt{6} \\
2\sqrt{3} & 0 & -2\sqrt{6} \\
2\sqrt{3} & -3\sqrt{2} & \sqrt{6}
\end{array}
\right) ,  \label{eq:C_3D_identical}
\end{equation}
that yields
\begin{eqnarray}
s_{1} &=&\sqrt{\frac{m}{3}}\left( q_{1}+q_{2}+q_{3}\right) ,  \nonumber \\
s_{2} &=&\sqrt{\frac{m}{2}}\left( q_{1}-q_{3}\right) ,  \nonumber \\
s_{3} &=&\sqrt{\frac{m}{6}}\left( 2q_{2}-q_{1}-q_{3}\right) .
\label{eq:s_i_3D_identical}
\end{eqnarray}
The occurrence of the eigenvalue $\lambda _{1}=0$ tells us that the center
of mass of the system moves freely at constant velocity. Note that the
variable $s_{1}$ is proportional to the coordinate of the center of mass $%
q_{CM}=\left( q_{1}+q_{2}+q_{3}\right) /3$.

\section{Conclusions}

\label{sec:conclusions}

The approach proposed by Hojman\cite{H23} is interesting in
itself. However, from a practical point of view the traditional
approach\cite{WDC55,C78} is more convenient because it appears to
be simpler. Although this approach is well known, we think that
its detailed application to particular simple examples may be of
pedagogical value to students of classical mechanics. For this
reason, we have applied it to all the models chosen by Hojman and
also to the case of equal frequencies that he avoided for
simplicity.

The problem discussed in appendix~\ref{sec:analogy} may also be of
pedagogical interest because it shows that two problems that students
commonly face in completely different courses (classical mechanics and
quantum mechanics) may be expressed in terms of identical mathematical
equations.

\appendix

\numberwithin{equation}{section}

\section{Analogy: Hermitian operator on a finite real vector space}

\label{sec:analogy}

In this appendix we discuss a well known mathematical problem that also
requires the simultaneous diagonalization of two matrices. Consider an
Hermitian operator $H$ defined on an $N$-dimensional real vector space
endowed with an inner product $\left\langle f\right. \left| g\right\rangle
=\left\langle g\right. \left| f\right\rangle $, for any $f$ and $g$ that
belong to the vector space. Such operator has a complete set of
eigenfunctions $\psi _{i}$ with eigenvalues $E_{i}$,
\begin{equation}
H\psi _{i}=E_{i}\psi _{i},\;i=1,2,\ldots ,N,  \label{eq:HPsi=Epsi}
\end{equation}
that we may choose to be orthonormal $\left\langle \psi _{i}\right. \left|
\psi _{j}\right\rangle =\delta _{ij}$. Suppose that $B=\left\{
f_{1},f_{2},\ldots ,f_{N}\right\} $ is a complete set of non-orthogonal
vectors $f_{i}$. Since each $\psi _{i}$ can be written as a linear
combination of the basis vectors
\begin{equation}
\psi _{i}=\sum_{j=1}^{N}c_{ji}f_{j},  \label{eq:psi_lim_comb}
\end{equation}
then we have
\begin{eqnarray}
\left\langle f_{k}\right| H\left| \psi _{i}\right\rangle
&=&\sum_{j=1}^{N}\left\langle f_{k}\right| H\left| f_{j}\right\rangle
c_{ji}=E_{i}\sum_{j=1}^{N}\left\langle f_{k}\right. \left|
f_{j}\right\rangle c_{ji}  \nonumber \\
&=&\sum_{j=1}^{N}\sum_{m=1}^{m}E_{m}\delta _{mi}\left\langle f_{k}\right.
\left| f_{j}\right\rangle c_{jm},
\end{eqnarray}
that can be written in matrix form as
\begin{equation}
\mathbf{HC=SCE},  \label{eq:HC=SCE}
\end{equation}
where $\left( \mathbf{H}\right) _{kj}=\left\langle f_{k}\right| H\left|
f_{j}\right\rangle $ , $\left( \mathbf{S}\right) _{kj}=\left\langle
f_{k}\right. \left| f_{j}\right\rangle $ and $\mathbf{E}=\left( E_{i}\delta
_{ij}\right) $. Therefore,
\begin{equation}
\mathbf{C}^{-1}\mathbf{S}^{-1}\mathbf{HC=E}.  \label{eq:CSHC=E}
\end{equation}
The orthonormality of the eigenfunctions leads to
\begin{equation}
\left\langle \psi _{i}\right. \left| \psi _{j}\right\rangle
=\sum_{k=1}^{N}\sum_{m=1}^{N}c_{ki}c_{mj}\left\langle f_{k}\right. \left|
f_{m}\right\rangle =\delta _{ij},
\end{equation}
that in matrix form reads
\begin{equation}
\mathbf{C}^{t}\mathbf{SC}=\mathbf{I.}  \label{eq:CSC=I}
\end{equation}
Note that the matrix $\mathbf{H}$ is symmetric and $\mathbf{S}$ is symmetric
and positive definite as in the case of the diagonalization of coupled
harmonic oscillators. In fact, equations (\ref{eq:CSC=I}) and (\ref
{eq:CSHC=E}) are identical, from a mathematical point of view, to equations (%
\ref{eq:CAC=I}) and (\ref{eq:CABC}), respectively. It is clear that the
problem posed by the diagonalization of two symmetric matrices is not
uncommon in mathematical physics.

\end{document}